\documentclass[aps,showpacs,showkeys,pre,twocolumn,superscriptaddress,floatfix]{revtex4}
 \usepackage{graphicx,bm,amssymb,amsmath,dcolumn,hyperref}
 \usepackage{multirow}
 \usepackage{enumerate}
 \usepackage{color}
\begin{document}
\newcommand{\be}{\begin{equation}}
\newcommand{\ee}{\end{equation}}
\newcommand{\half}{\frac{1}{2}}
\newcommand{\ith}{^{(i)}}
\newcommand{\im}{^{(i-1)}}
\newcommand{\gae}
{\,\hbox{\lower0.5ex\hbox{$\sim$}\llap{\raise0.5ex\hbox{$>$}}}\,}
\newcommand{\lae}
{\,\hbox{\lower0.5ex\hbox{$\sim$}\llap{\raise0.5ex\hbox{$<$}}}\,}

\definecolor{blue}{rgb}{0,0,1}
\definecolor{red}{rgb}{1,0,0}
\definecolor{green}{rgb}{0,1,0}
\newcommand{\blue}[1]{\textcolor{blue}{#1}}
\newcommand{\red}[1]{\textcolor{red}{#1}}
\newcommand{\green}[1]{\textcolor{green}{#1}}

\newcommand{\scrA}{{\mathcal A}} 
\newcommand{\scrL}{{\mathcal L}} 
\newcommand{\scrN}{{\mathcal N}} 
\newcommand{\scrS}{{\mathcal S}} 
\newcommand{\scrs}{{\mathcal s}} 
\newcommand{\scrP}{{\mathcal P}} 
\newcommand{\scrO}{{\mathcal O}} 
\newcommand{\scrR}{{\mathcal R}} 
\newcommand{\scrC}{{\mathcal C}} 

\newcommand{\dm}{d_{\rm min}}

\newcommand{\RbondFig}{\includegraphics[scale=0.70]{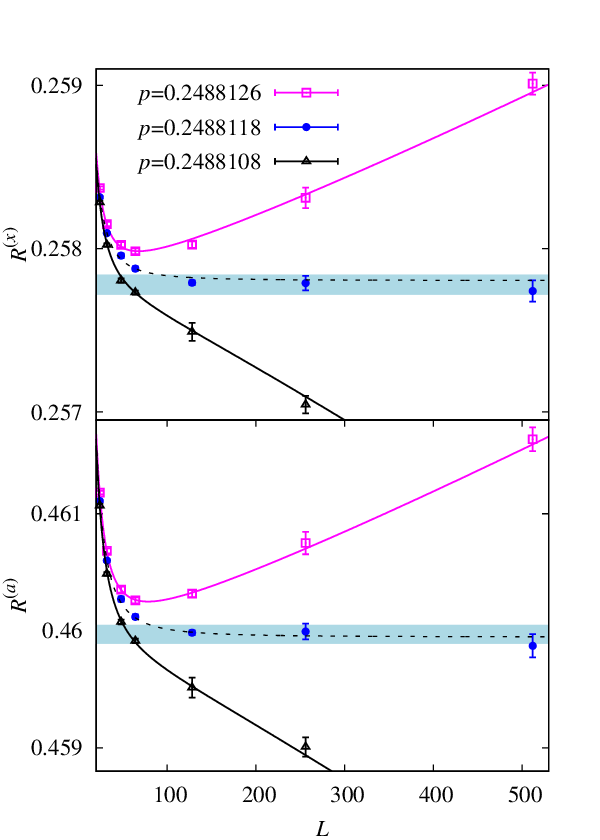}}
\newcommand{\yiFiga}{\includegraphics[scale=0.65]{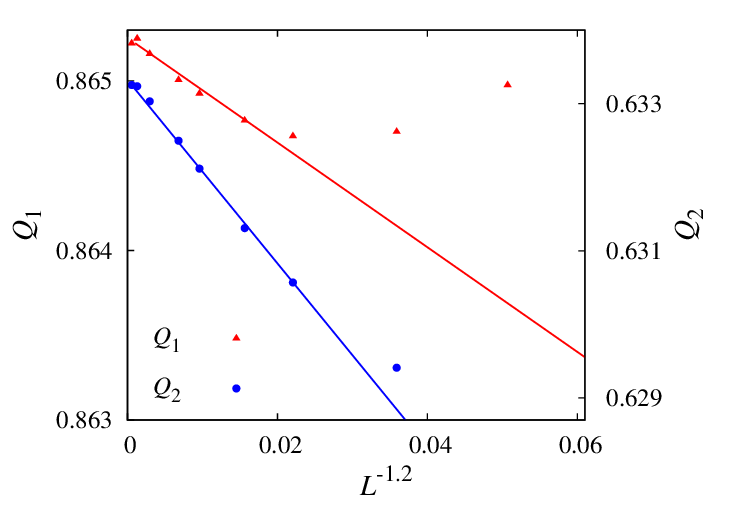}}
\newcommand{\yiFigb}{\includegraphics[scale=0.65]{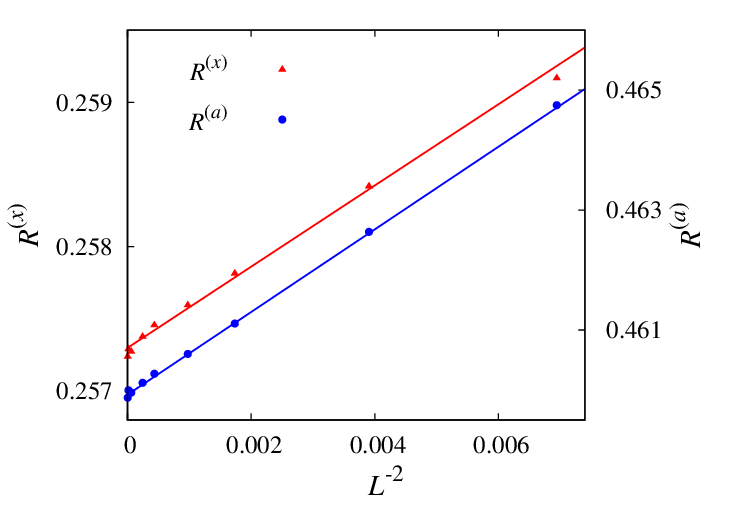}}
\newcommand{\RsiteFig}{\includegraphics[scale=0.75]{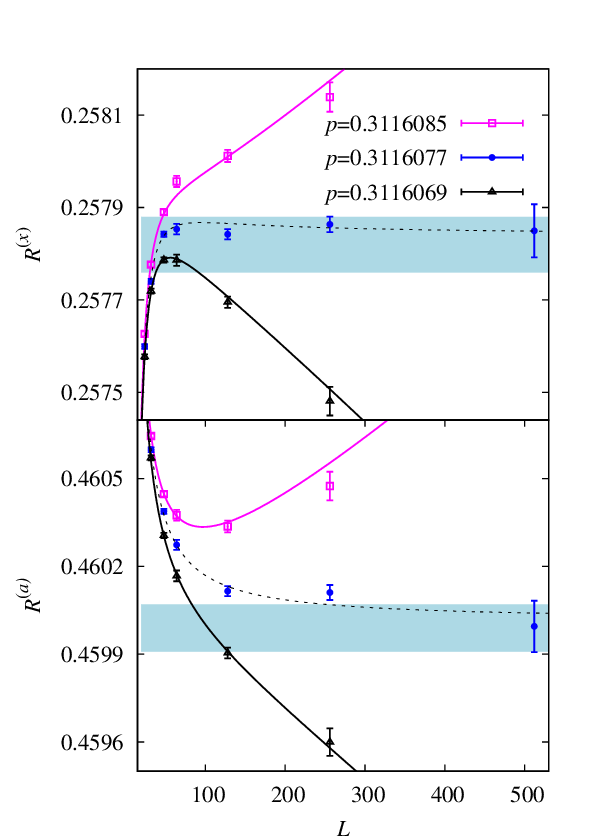}}
\newcommand{\ytFig}{\includegraphics[scale=0.70]{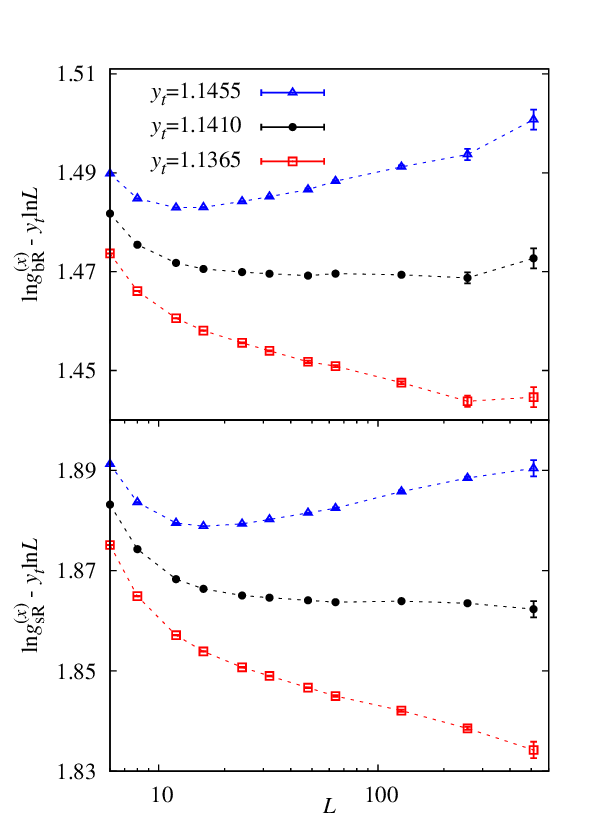}}
\newcommand{\yhFig}{\includegraphics[scale=0.70]{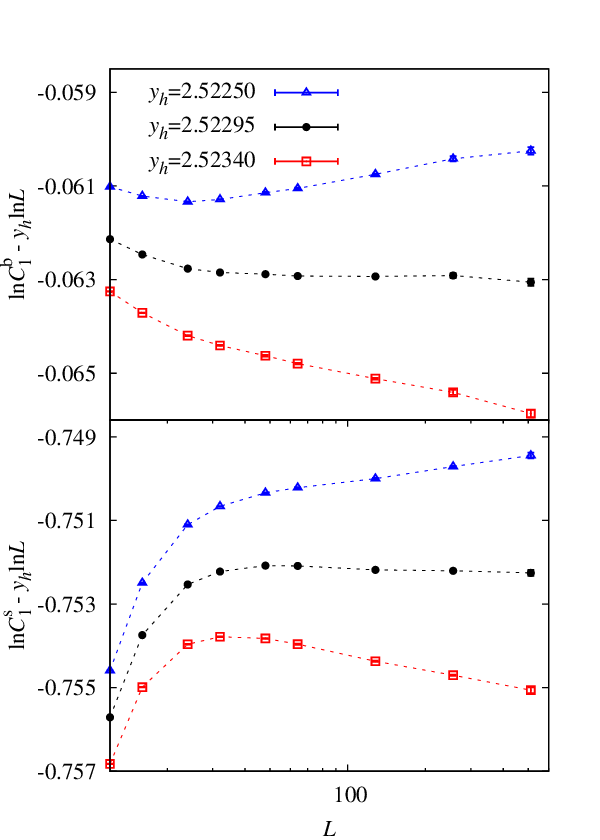}}
\newcommand{\dminFig}{\includegraphics[scale=0.65]{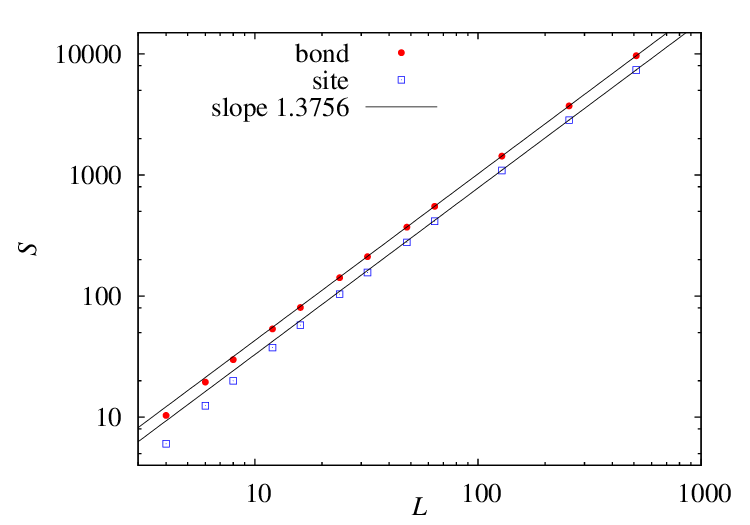}}

\title{Bond and Site Percolation in Three Dimensions}
\date{\today} 
\author{Junfeng Wang}
\affiliation{Hefei National Laboratory for Physical Sciences at Microscale and Department of Modern Physics, University of Science and Technology of China, Hefei, Anhui 230026, China}
\author{Zongzheng Zhou}
\affiliation{Hefei National Laboratory for Physical Sciences at Microscale and Department of Modern Physics, University of Science and Technology of China, Hefei, Anhui 230026, China}
\affiliation{School of Mathematical Sciences, Monash University, Clayton, Victoria~3800, Australia}
\author{Wei Zhang}
\affiliation{Department of Physics, Jinan University, Guangzhou 510632, China}
\author{Timothy M. Garoni}
\email{tim.garoni@monash.edu}
\affiliation{School of Mathematical Sciences, Monash University, Clayton, Victoria~3800, Australia}
\author{Youjin Deng}
\email{yjdeng@ustc.edu.cn}
\affiliation{Hefei National Laboratory for Physical Sciences at Microscale and Department of Modern Physics, University of Science and Technology of China, Hefei, Anhui 230026, China}

\begin{abstract}
  We simulate the bond and site percolation models on a simple-cubic lattice with linear sizes up to $L=512$,
  and estimate the percolation thresholds to be $p_c ({\rm bond})=0.248\,811\,82(10)$ and $p_c ({\rm site})=0.311\,607\,7(2)$.
  By performing extensive simulations at these estimated critical points, we then estimate the critical exponents
  $1/\nu =1.141\,0(15)$, $\beta/\nu=0.477\,05(15)$, the leading correction exponent $y_i =-1.2(2)$, and the shortest-path exponent $\dm=1.375\,6(3)$.
  Various universal amplitudes are also obtained, including wrapping probabilities, 
  ratios associated with the cluster-size distribution,
  and the excess cluster number.
  We observe that the leading finite-size corrections in certain wrapping probabilities are governed by
  an exponent $\approx -2$, rather than $y_i \approx -1.2$.
\end{abstract}
\maketitle 

\section{Introduction}
 Percolation~\cite{BroadbentHarmmersley57} is a cornerstone of the theory of critical phenomena~\cite{StaufferAharony94},
and a central topic in probability~\cite{Grimmett99,BollobasRiordan09}.
 In two dimensions, Coulomb gas arguments~\cite{Nienhuis87} and
conformal field theory~\cite{Cardy87} predict the exact values of the bulk critical exponents $\beta =5/36$ and $\nu=4/3$, 
which have been confirmed rigorously in the specific case of triangular-lattice site percolation~\cite{SmirnovWerner01}.
Exact values of the percolation thresholds $p_c$ on several two-dimensional lattices are also known~\cite{Essam72}. 
In particular, it is known rigorously~\cite{Kesten80} that $p_c=1/2$ for bond percolation on the square lattice.
For all $d$ greater than or equal to the upper critical dimension~\cite{Toulouse74} of $d_c=6$, the mean-field values for the exponents $\beta=1$ and
$d \nu=3$ are believed to hold; this has been proved rigorously ~\cite{AizenmanNewman84,HaraSlade90} for $d\ge19$.

For dimensions $2 < d < 6$ by contrast, no exact values for either the critical exponents or the percolation thresholds are known.
Significant effort has therefore been expended on obtaining ever more accurate estimates, especially in three dimensions.

In addition to percolation thresholds and critical exponents, crossing probabilities \cite{LanglandsPichetPouliotSaintAubin92,Cardy92}
also play an important role in studies of percolation.
For lattices drawn on a torus, the analogous quantities are wrapping probabilities~\cite{LanglandsPouliotSaintAubin94}, and in two dimensions their values can be determined exactly~\cite{Pinson94}.
The three-dimensional case~\cite{MartinsPlascak03} has been far less studied however.
Precise estimation of wrapping probabilities on the simple-cubic lattice represents one of the central undertakings of the current work.

In addition to their intrinsic importance, wrapping probabilities have proved to be an effective practical means of estimating percolation thresholds~\cite{NewmanZiff00,FengDengBlote08}.
Using Monte Carlo (MC) simulations and performing a careful finite-size scaling analysis of various wrapping probabilities in the neighborhood of the transition, 
we obtain very accurate estimates of $p_c$ for both site and bond percolation. 
We observe numerically that the leading finite-size corrections for certain wrapping probabilities appear to be governed by an exponent $\approx-2$, 
rather than by the leading irrelevant exponent $y_i\approx-1.2$.

We then estimate the thermal exponent $y_t=1/\nu$ by fixing $p$ to our best estimate of $p_c$, and
studying the divergence with linear size $L$ of the derivative of the wrapping probability, which is proportional to the covariance of its indicator with the number of bonds.
We find this procedure for estimating $y_t$ preferable to methods in which $y_t$ is estimated by studying 
how quantities behave in a neighborhood of $p$ values around $p_c$. In particular, we believe the current method produces more reliable error estimates.

The remainder of this paper is organized as follows. The simulation method and the sampled quantities 
are discussed in Sec.~\ref{sampled quantities}.
The results for the wrapping probabilities and thresholds are given in Sec.~\ref{estimating pc}.
Critical exponents and the excess cluster number are discussed in Sec.~\ref{results at pc}.
We then finally conclude with a discussion in Sec.~\ref{discussion}.

 \begin{table*}
 \begin{center}
 \caption{Fits of the wrapping probabilities $R^{(x)}$, $R^{(a)}$, and $R^{(3)}$, and the ratios $Q_1$ and $Q_2$ for bond percolation.
         We did not obtain stable fits with $y_i$ free for $R^{(3)}$.}
 \scalebox{1.0}{
 \begin{tabular}[t]{|l|l|l|l|l|l|l|l|l|l|}
 \hline 
                               & $L_{\rm min}$ & $\chi^2/$DF& $p_c$     & $y_t$    & $\scrO_c$       & $q_1$      & $b_1$   & $y_i$     & $b_2$\\
 \hline
 {\multirow{6}{*}{$Q_1$}}      & 16            & 53/40  & 0.248\,812\,03(5) & 1.16(1)  & 0.865\,37(1)    & $-0.36(1)$ &$-0.0423(5)$& $-1.2$  & ~~0.341(5)\\
                               & 24            & 33/33  & 0.248\,811\,98(6) & 1.16(2)  & 0.865\,35(2)    & $-0.31(2)$ &$-0.040(2)$ & $-1.2$  & ~~0.31(2) \\
                               & 32            & 28/26  & 0.248\,811\,93(7) & 1.19(3)  & 0.865\,33(2)    & $-0.31(3)$ &$-0.036(3)$ & $-1.2$  & ~~0.25(5) \\
 \cline{2-10}
                               & 16            & 44/39  & 0.248\,811\,84(8)  & 1.16(1)  & 0.865\,39(3)   & $-0.36(1)$ &$-0.10(4) $ & $-1.34(9)$ &~~$0.50(8)$  \\
                               & 24            & 31/32  & 0.248\,811\,88(9)  & 1.19(2)  & 0.865\,29(4)   & $-0.32(3)$ &$-0.10(8) $ & $-1.3(2)$  &~~$0.5(2)$   \\
                               & 32            & 28/25  & 0.248\,811\,96(14) & 1.19(3)  & 0.865\,4(2)    & $-0.31(3)$ &$-0.02(4) $ & $-1.0(5)$  &~~$0.2(3)$     \\
 \hline
 {\multirow{6}{*}{$Q_2$}}      & 32            & 28/25  & 0.248\,811\,20(5) & 1.17(2)  & 0.633\,58(3)    & $-0.80(5)$ &$-0.104(4)$ & $-1.2$  & ~~0.05(7) \\
                               & 48            & 16/18  & 0.248\,811\,95(6) & 1.14(2)  & 0.633\,50(3)    & $-0.89(8)$ &$-0.088(9)$ & $-1.2$  & $-0.3(2)$  \\
                               & 64            & 10/11  & 0.248\,811\,84(11)& 1.12(3)  & 0.633\,4(2)     & $-1.0(2)$  &$-0.05(4)$  & $-1.2$  & $-1(1)$  \\
 \cline{2-10}
                               & 32            & 28/26  & 0.248\,812\,02(6) & 1.17(2)  & 0.633\,58(5)    & $-0.80(5)$ &$-0.097(8)$ & $-1.08(3)$  & ~~~~- \\
                               & 48            & 16/19  & 0.248\,811\,93(7) & 1.14(2)  & 0.633\,46(7)    & $-0.89(8)$ &$-0.15(4) $ & $-1.22(7)$  & ~~~~- \\
                               & 64            & 10/12  & 0.248\,811\,82(11)& 1.12(3)  & 0.633\,3(2)     & $-1.0(2)$  &$-0.5(6) $  & $-1.5(4)$   & ~~~~- \\
 \hline
 {\multirow{6}{*}{$R^{(x)}$}}  & 16            & 41/37  & 0.248\,811\,81(4) & 1.143(7) & 0.257\,77(2)  & $-1.22(3)$ &~~0.005(2) & $-1.2$    & $-0.23(1)$ \\
                               & 24            & 30/31  & 0.248\,811\,83(4) & 1.15(2)  & 0.257\,78(3)  & $-1.22(6)$ &~~0.003(3) & $-1.2$    & $-0.26(4)$ \\
                               & 32            & 25/24  & 0.248\,811\,82(6) & 1.15(2)  & 0.257\,76(5)  & $-1.20(8)$ &~~0.006(7) & $-1.2$    & $-0.20(10)$\\ 
 \cline{2-10}
                               & 16            & 41/37  & 0.248\,811\,82(4) & 1.144(7) & 0.257\,79(2)  & $-1.22(3)$ &~~0.18(2)  & $-1.83(4)$ & ~~~~-  \\
                               & 24            & 31/31  & 0.248\,811\,84(4) & 1.15(2)  & 0.257\,79(2)  & $-1.22(6)$ &~~0.22(8)  & $-1.9(2)$  & ~~~~-  \\
                               & 32            & 25/24  & 0.248\,811\,82(6) & 1.15(2)  & 0.257\,77(4)  & $-1.20(8)$ &~~0.1(1)   & $-1.7(3)$  & ~~~~-  \\
 \hline
 {\multirow{6}{*}{$R^{(a)}$}}  & 16            & 40/39  & 0.248\,811\,82(4) & 1.149(7) & 0.459\,99(3)  & $-1.65(4)$ &~~0.004(2)& $-1.2$  & ~~$0.73(2)$ \\
                               & 24            & 25/32  & 0.248\,811\,82(5) & 1.14(2)  & 0.459\,97(5)  & $-1.74(9)$ &~~0.003(4)& $-1.2$  & ~~$0.72(6)$\\
                               & 32            & 22/25  & 0.248\,811\,83(6) & 1.14(2)  & 0.459\,98(7)  & $-1.7(2) $ &~~0.005(9)& $-1.2$  & ~~$0.7(2)$   \\
 \cline{2-10}
                               & 16            & 40/39  & 0.248\,811\,82(4) & 1.149(7) & 0.459\,97(2)  & $-1.65(4)$ &~~0.81(6) & $-2.06(3)$ & ~~~~- \\
                               & 24            & 25/32  & 0.248\,811\,82(4) & 1.14(2)  & 0.459\,95(3)  & $-1.74(9)$ &~~0.8(2)  & $-2.05(8)$ & ~~~~- \\
                               & 32            & 22/25  & 0.248\,811\,82(5) & 1.14(2)  & 0.459\,96(5)  & $-1.74(2)$ &~~1.0(9)  & $-2.1(3)$  & ~~~~- \\
 \hline
 {\multirow{3}{*}{$R^{(3)}$}}  & 16            & 44/38  & 0.248\,811\,85(6 )& 1.14(1)  & 0.080\,41(2)  & $-0.66(2)$ &~~0.010(1)& $-1.2$  & $-0.076(8)$  \\
                               & 24            & 35/31  & 0.248\,811\,91(6 )& 1.15(2)  & 0.080\,43(3)  & $-0.63(5)$ &~~0.007(3)& $-1.2$  & $-0.04(3)$   \\
                               & 32            & 23/24  & 0.248\,811\,85(8 )& 1.17(3)  & 0.080\,39(4)  & $-0.59(5)$ &~~0.014(6)& $-1.2$  & $-0.15(9)$   \\
 \hline
 \end{tabular}
}
 \end{center}
 \label{Tab:fit-bond}
 \end{table*}

\section{Sampled quantities}
\label{sampled quantities}
We study bond and site percolation on a periodic $L\times L\times L$ simple-cubic lattice with linear system sizes $L=8$, 12, 16, 24, 32, 48, 64, 128, 256, and 512.
For each system size, we produced at least $2.5\times 10^7$ independent samples.
Each independent bond (site) configuration is generated by independently occupying each bond (site) with probability $p$.
The clusters in each configuration are identified using breadth-first search. 
The number of sites in each cluster defines its size.

 We sampled the following observables in our simulations:
 \begin{enumerate}[(a)]
 \item The number of occupied bonds $\scrN_b$ for bond percolation, and the number of occupied sites $\scrN_s$ for site percolation.
 \item The number of clusters $\scrN_c$.
 \item The size $\scrC_1$ of the largest cluster.
 \item The cluster-size moments $\scrS_m  = \sum_{C}|C|^m$ with $m=0,2,4$. The sum runs over all clusters $C$, and $\scrS_0$ is simply the number of clusters.
 \item An observable $\scrS :=\max\limits_{C}\,\max\limits_{y\in C}\,d(x_C,y)$ used to determine the shortest-path exponent.
   Here $d(x,y)$ denotes the graph distance from site $x$ to site $y$, and $x_C$ is the vertex in cluster $C$ with the smallest vertex label, according to some fixed (but arbitrary) vertex labeling.
 \item The indicators $\scrR^{(x)}$, $\scrR^{(y)}$, and $\scrR^{(z)}$, for the event that a cluster wraps around the lattice in the $x$, $y$, or $z$ direction, respectively.
 \end{enumerate}

 From these observables we calculated the following quantities:
 \begin{enumerate}[(i)]
 \item The mean size of the largest cluster $C_1 = \langle \scrC_1 \rangle$, which at $p_c$ scales as $C_1\sim L^{y_h}$ with
$y_h = d_f = d - \beta/\nu$, where $d_f$ is the fractal dimension.
 \item The cluster density $\rho = \langle \scrN_c \rangle/L^d$.
 \item The mean size of the cluster at the origin, $\chi=\langle \scrS_2 \rangle/L^d$, which at $p_c$ scales as $\chi \sim L^{2y_h-d}$.
 \item The dimensionless ratios
   \begin{equation}
     Q_1  = \frac{\langle \scrC_1\rangle^2}{\langle {\scrC_1}^2\rangle}\;,\;\;\; Q_2  = \frac{\langle {\scrS_2}^2\rangle}{\langle 3{\scrS_2}^2 - 2\scrS_4\rangle}\;.
     \label{eq:R}
   \end{equation}
 \item The shortest-path length $S=\langle \scrS \rangle$, which at $p_c$ scales as $S \sim L^{\dm}$ with $\dm$ the shortest-path fractal dimension.
 \item The wrapping probabilities
 \begin{equation}
   \begin{aligned}
     R^{(x)} = & \langle \scrR^{(x)} \rangle = \langle \scrR^{(y)} \rangle = \langle \scrR^{(z)} \rangle \;, \\
     R^{(a)} = & 1 - \langle (1-\scrR^{(x)})(1-\scrR^{(y)})(1-\scrR^{(z)})\rangle \;, \\
     R^{(3)} = & \langle \scrR^{(x)}\scrR^{(y)}\scrR^{(z)}\rangle\;.
   \end{aligned}
 \end{equation}
 Here $R^{(x)}$ gives the probability that a winding exists in the $x$ direction, $R^{(a)}$ gives the probability that a winding exists in at least one of the three possible directions,
 and $R^{(3)}$ gives the probability that windings simultaneously exist in all three possible directions.
 Near $p_c$, we expect each of these wrapping probabilities to behave as $\sim f((p-p_c)L^{y_t})$, where $f$ is a scaling function.
 \item The covariance of $\scrR^{(x)}$ and $\scrN_b$
 \begin{equation}
 \begin{aligned}
   g^{(x)}_{bR} &= \langle \scrR^{(x)} \scrN_b\rangle - \langle \scrR^{(x)}\rangle \langle \scrN_b \rangle \\
              &= p(1-p)\frac{\partial R^{(x)}}{\partial p}\;.
 \end{aligned}
 \label{eq:g}
 \end{equation}
 At $p_c$, we expect $g^{(x)}_{bR}\sim L^{y_t}$.
 An analogous definition of $g^{(x)}_{sR}$, with $\scrN_b$ being replaced with $\scrN_s$, was used for site percolation.
 \end{enumerate}

 To derive \eqref{eq:g}, one can explicitly differentiate $\langle \scrR^{(x)}\rangle$ with respect to $p$,
 and use the fact that $\langle \scrN_b \rangle = p|E|$ where $|E|$ is the total number of edges on the lattice.

 The complete set of data for all observables, for both bond and site percolation, 
 is available as Supplemental Material \cite{Supplemental}.
% contained as a file \texttt{percolation.tar.gz} in the preprint version of this paper at  \href{http://www.arxiv.org}{arXiv.org}
%~\footnote{The tar file unpacks to make a directory \texttt{percolation} with subdirectories \texttt{bond} and \texttt{site}.
%   Individual files have names like \texttt{C1\_bond.txt} or \texttt{CHI\_site.txt} and have four fields on each line: $L$, $p$, value and error bar.}.

 \begin{figure}
 \RbondFig
 \centering
 \caption{
   Plots of $R^{(x)}(p,L)$ (top) and $R^{(a)}(p,L)$ (bottom) vs $L$ for fixed values of $p$, for bond percolation.
   In both cases, the curves correspond to our preferred fit of the MC data for $R(p,L)$ by the ansatz~(\ref{eq:pc}); 
   the dashed curve corresponds to setting $p=0.248\,811\,82$.
   The shaded blue strips indicate an interval of $1\sigma$ above and below the estimates $R^{(x)}_c = 0.257\,78(6)$ and $R^{(a)}_c = 0.459\,97(8)$.
 }
 \label{Fig:R-pc}
 \end{figure}

 \begin{figure}
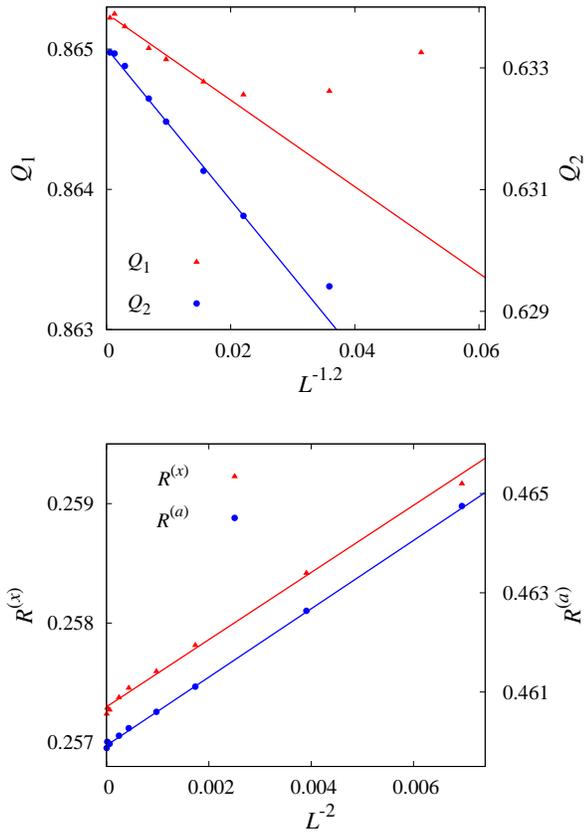

 \centering
 \yiFiga \\ 
 \yiFigb
 \caption{
  Plots of $Q_1$ and $Q_2$ vs $L^{-1.2}$ (top), 
  and $R^{(x)}$ and $R^{(a)}$ vs $L^{-2}$ (bottom), with $p=0.248\,811\,8$, for bond percolation.
   The solid lines are simply to guide the eye.
 }
 \label{Fig:yi}
 \end{figure}

\section{Estimating $p_c$}
\label{estimating pc}
\subsection{Bond percolation}

 We estimate the thresholds of bond and site percolation by studying the finite-size scaling of the 
 wrapping probabilities $R^{(x)}$, $R^{(a)}$, and $R^{(3)}$, and the dimensionless ratios $Q_1$ and $Q_2$.
 Around $p_c$, we perform least-squares fits of the MC data for these quantities by the ansatz
 \be
 \scrO(\epsilon,L)  = \scrO_c + \sum_{k=1}^{2}q_k\epsilon^kL^{ky_t} + b_1L^{y_i} + b_2L^{-2}\;\;,
 \label{eq:pc}
 \ee
 where $\epsilon = p_c - p$, $\scrO_c$ is a universal constant, and $y_i$ is the leading correction exponent.
 We perform fits with both $b_1$ and $b_2$ free, as well as fits with $b_2$ being set identically to zero. 
 By performing fits with $y_i$ free we estimate that $y_i=-1.2(2)$. 
 We also perform fits with $y_i$ fixed to $y_i=-1.2$.

As a precaution against correction-to-scaling terms that we have neglected in our chosen ansatz, 
we impose a lower cutoff $L\ge L_{\min}$ on the data points admitted in the fit, 
and we systematically study the effect on the $\chi^2$ value of increasing $L_{\min}$. 
In general, our preferred fit for any given ansatz corresponds to the smallest $L_{\min}$ 
for which $\chi^2$ divided by the number of degrees of freedom (DFs) is $O(1)$, 
and for which subsequent increases in $L_{\min}$ do not cause $\chi^2$ to drop by much more than one unit per degree of freedom.

 \begin{table*}
 \begin{center}
 \caption{Fits of the wrapping probabilities $R^{(x)}$, $R^{(a)}$, and $R^{(3)}$, and the ratios $Q_1$ and $Q_2$ for site percolation.
 For $R^{(x)}$ we obtain unstable results when $y_i$ is free.}
 \scalebox{1.0}{
 \begin{tabular}[t]{|l|l|l|l|l|l|l|l|l|l|}
 \hline
                               & $L_{\rm min}$ & $\chi^2$/DF & $p_c$  & $y_t$    & $\scrO_c$     & $q_1$      & $b_1$           & $y_i$    & $b_2$\\
 \hline
  {\multirow{6}{*}{$Q_1$}}     & 32            & 19/16  & 0.311\,606\,9(2)  & 1.14(2)  & 0.865\,05(2)    & $-0.22(2)$ &~~$0.062(2) $ & $-1.2$  &~~0(3) \\
                               & 48            & 11/11  & 0.311\,607\,0(2)  & 1.11(3)  & 0.865\,09(3)    & $-0.25(3)$ &~~$0.054(6) $ & $-1.2$  &~~0.2(2) \\
                               & 64            & ~~3/6    & 0.311\,607\,7(3)  & 1.12(6)  & 0.865\,26(7)  & $-0.24(6)$ &~~$0.01(2)  $ & $-1.2$  &~~1.4(5) \\
 \cline{2-10}
                               & 32            & 19/16  & 0.311\,606\,9(2)  & 1.15(2)  & 0.865\,06(3)    & $-0.22(2)$ &~~$0.063(4) $ & $-1.11(2)$ & ~~~~~~-\\
                               & 48            & 10/11  & 0.311\,607\,1(2)  & 1.11(3)  & 0.865\,12(4)    & $-0.25(3)$ &~~$0.09(2)  $ & $-1.22(7)$ & ~~~~~~-\\
                               & 64            & ~~3/6    & 0.311\,607\,7(3)  & 1.12(6)  & 0.865\,27(5)  & $-0.24(6)$ &~~$0.9(10)  $ & $-1.8(3) $ & ~~~~~~-\\
 \hline
  {\multirow{3}{*}{$Q_2$}}     & 64            &~~3/6    & 0.311\,607\,6(2)  & 1.12(4)  & 0.633\,3(1)     & $-0.56(9)$ &~~$0.02(3)  $ & $-1.2$  &~~5.1(7) \\
 \cline{2-10}                  & 48            &13/11   & 0.311\,607\,2(1)  & 1.14(2)  & 0.633\,06(4)    & $-0.52(4)$ &~~$0.9(1)  $ & $-1.52(3)$ &~~~~~~-   \\
                               & 64            &~~2/6     & 0.311\,607\,6(2)  & 1.12(4)  & 0.633\,29(8)    & $-0.56(9)$ &~~$ 4(2)    $ & $-1.9(2) $ &~~~~~~-   \\
 \hline
  {\multirow{3}{*}{$R^{(x)}$}} &16          & 42/39  & 0.311\,607\,85(5) & 1.13(1)  & 0.257\,89(2)  & $-0.76(4)$ &~~$0.004(1) $   & $-1.2$ & $-0.22(1)$  \\
                               &24          & 30/31  & 0.311\,607\,74(6) & 1.14(2)  & 0.257\,84(3)  & $-0.75(5)$ &~~$0.009(2)  $   & $-1.2$ & $-0.29(3)$   \\
                               &32          & 24/24  & 0.311\,607\,66(7) & 1.14(2)  & 0.257\,80(3)  & $-0.73(5)$ &~~$0.015(4)$     & $-1.2$ & $-0.39(6)$   \\
 \hline
  {\multirow{6}{*}{$R^{(a)}$}} &16          & 39/40  & 0.311\,607\,70(5) & 1.12(2)  & 0.460\,02(2)  & $-1.09(6)$ &~~0.023(2)       &$-1.2 $  &~~$0.08(2)$    \\
                               &24          & 25/32  & 0.311\,607\,67(7) & 1.13(2)  & 0.459\,99(4)  & $-1.05(6)$ &~~0.025(3)       &$-1.2 $  &~~$0.05(4)$    \\
                               &32          & 19/24  & 0.311\,607\,65(8) & 1.13(2)  & 0.459\,98(5)  & $-1.06(7)$ &~~0.027(6)       &$-1.2 $  &~~$0.02(9)$    \\
 \cline{2-10}
                               &16          & 36/40  & 0.311\,607\,75(6) & 1.12(2)  & 0.460\,06(3)  & $-1.09(6)$ &~~0.055(5)      &$-1.33(4)$ &~~~~~~-         \\
                               &24          & 25/32  & 0.311\,607\,68(8) & 1.13(2)  & 0.460\,01(5)  & $-1.05(7)$ &~~0.039(9)      &$-1.21(9)$ &~~~~~~-         \\
                               &32          & 19/24  & 0.311\,607\,65(9) & 1.13(2)  & 0.459\,99(7)  & $-1.06(7)$ &~~0.03(2)       &$-1.1(2)$  &~~~~~~-          \\
\hline
  {\multirow{6}{*}{$R^{(3)}$}} &16          & 50/38  & 0.311\,608\,01(8) & 1.14(2)  & 0.080\,55(1)  & $-0.38(3)$ & $-0.010(8)$   &$-1.2$  & $-0.30(1)$ \\
                               &24          & 27/30  & 0.311\,607\,79(9) & 1.14(2)  & 0.080\,49(2)  & $-0.39(4)$ & $-0.004(2)$   &$-1.2$  & $-0.38(3)$  \\
                               &32          & 18/23  & 0.311\,607\,65(11)& 1.15(3)  & 0.080\,45(3)  & $-0.38(4)$ & $-0.002(3)$   &$-1.2$  & $-0.47(5)$  \\
\cline{2-10}
                               &16          & 40/38  & 0.311\,607\,89(7) & 1.15(2)  & 0.080\,510(9) & $-0.38(3)$ & $-0.21(1) $   &$-1.77(2)$ &~~~~~~-     \\
                               &24          & 26/30  & 0.311\,607\,77(8) & 1.14(2)  & 0.080\,48(2)  & $-0.39(4)$ & $-0.30(5) $   &$-1.88(5)$ &~~~~~~-    \\
                               &32          & 18/23  & 0.311\,607\,66(10)& 1.15(3)  & 0.080\,46(2)  & $-0.38(4)$ & $-0.6(2)  $   &$-2.1(2)$  &~~~~~~-     \\
\hline
 \end{tabular}
}
 \label{Tab:fit-site}
 \end{center}
 \end{table*}

 Table~\ref{Tab:fit-bond} summarizes the results of these fits.
 From the fits, we can see that the finite-size corrections of $Q_1$ and $Q_2$ are dominated by the exponent $y_i\approx -1.2$.
 From $Q_1$ and $Q_2$, we estimate $p_c=0.248\,811\,9(3)$, and their universal critical values $Q_{1,c}=0.865\,4(2)$ and $Q_{2,c}=0.633\,5(2)$.

 For $R^{(x)}$ and $R^{(a)}$, fixing $y_i=-1.2$ and including both the $b_1$ and $b_2$ terms we find that $b_1$ is consistent with zero, while $b_2$ is clearly nonzero.
 Furthermore, if we set $b_2=0$ and leave $y_i$ free, we find $y_i\approx -2$.
 This suggests that either the amplitudes of the leading corrections of $R^{(x)}$ and $R^{(a)}$ vanish identically, or at least that they are sufficiently small that they cannot be detected from our data.
 Due to these weak finite-size corrections, the values of $p_c$ fitted from $R^{(x)}$ and $R^{(a)}$ 
are much more stable than those obtained from $Q_1$ and $Q_2$.
 From $R^{(x)}$ and $R^{(a)}$, we estimate $p_c=0.248\,811\,82(10)$.
 For $R^{(3)}$, we report only the fits with corrections $b_1L^{-1.2} + b_2L^{-2}$.
 If $y_i$ is left free the fits become unstable, regardless of whether the $b_2L^{-2}$ term is included.
 From $R^{(3)}$, we estimate $p_c = 0.248\,811\,85(15)$ which is consistent with the value obtained from $R^{(x)}$ and $R^{(a)}$.
 From these fits, we estimate the universal wrapping probabilities to be $R^{(x)}_{c} = 0.257\,78(6)$, 
 $R^{(a)}_{c} = 0.459\,97(8)$ and $R^{(3)}_{c} = 0.080\,41(8)$.

 In Fig.~\ref{Fig:R-pc}, we illustrate our estimate of $p_c$ by plotting $R^{(x)}$ and $R^{(a)}$ vs $L$.
 Precisely at $p=p_c$, as $L\to\infty$ the data should tend to a horizontal line, whereas the data with $p\neq p_c$ will bend upward or downward.
 Figure~\ref{Fig:R-pc} shows that our estimate of $p_c$ lies slightly below the central value $0.248\,812\,6$ reported in~\cite{LorenzZiff98a}.

 In Fig.~\ref{Fig:yi}, we plot the data at $p=0.248\,811\,8$ for $R^{(x)}$ and $R^{(a)}$ vs $L^{-2}$, and for $Q_1$ and $Q_2$ vs $L^{-1.2}$.
 The figure strongly suggests that the correction $L^{-1.2}$ dominates in $Q_1$ and $Q_2$, but vanishes (or is very weak) in $R^{(x)}$ and $R^{(a)}$.
 
\subsection{Site percolation}
 For site percolation, we again estimate $p_c$ by fitting $Q_1$ and $Q_2$, $R^{(x)}$, $R^{(a)}$, and $R^{(3)}$ with Eq.~\eqref{eq:pc}.
 The fitting procedure is similar to that of bond percolation, and the results are summarized in Table~\ref{Tab:fit-site}.
 From the table, we can see that the fits of $Q_1$ and $Q_2$ are less stable for site percolation than for bond percolation.
 The ratio of $\chi^2$ per $\mathrm{DF}$ remains large until $L_{\rm min}\ge 32$ for $Q_1$ and $L_{\rm min}\ge48$ for $Q_2$,
 and the resulting estimates of $p_c$ range from $0.311\,606\,9(2)$ to $0.311\,607\,7(3)$.

 The fits of the wrapping probabilities are better behaved, as was the case for bond percolation.
 For $R^{(3)}$, fixing $y_i=-1.2$ and including both the $b_1$ and $b_2$ terms, we find that $b_1$ is consistent with zero, while $b_2$ is clearly nonzero.
 Furthermore, if we set $b_2=0$ and leave $y_i$ free, we find $y_i\approx -2$.
 This suggests that the amplitude of the leading correction of $R^{(3)}$ is smaller than the resolution of our fits, and might possibly be zero.
 The fits of the $R^{(a)}$ data, however, quite clearly indicate the presence of the $b_1L^{-1.2}$ term.
 For $R^{(x)}$, we report only the fits with corrections $b_1L^{-1.2} + b_2L^{-2}$;
 if $y_i$ is left free the fits become unstable, regardless of whether the $b_2L^{-2}$ term is included.
 As for $R^{(a)}$, the amplitude $b_1$ appears to take a nonzero value.
 These observations suggest that the leading correction $L^{-1.2}$ does not generically vanish for all wrapping probabilities, but rather that the amplitudes in some cases are smaller than the resolution of our
 simulations.

\begin{figure}
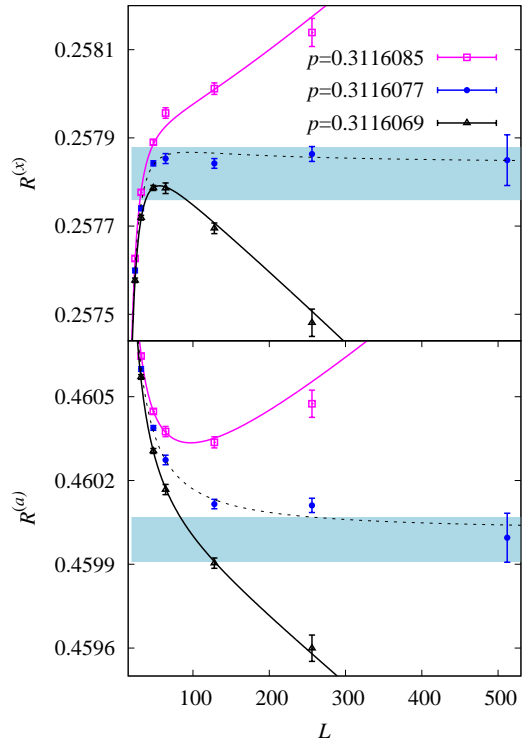

 \RsiteFig
 \centering
 \caption{
   Plots of $R^{(x)}(p,L)$ (top) and $R^{(a)}(p,L)$ (bottom) vs $L$ for fixed values of $p$, for site percolation.
   In both cases, the curves correspond to our preferred fit of the MC data for $R(p,L)$ by ansatz~(\ref{eq:pc}); 
    the dashed curve corresponds to setting $p=0.311\,607\,7$.
   The shaded blue strips indicate an interval of $1\sigma$ above and below the estimates $R^{(x)}_c = 0.257\,82(6)$ and $R^{(a)}_c = 0.459\,99(8)$.
 }
 \label{Fig:site-R}
 \end{figure}

 Comparing the various fits, we estimate $p_c=0.311\,607\,7(2)$ for site percolation, which is consistent with
 the previous result $0.311\,607\,7(4)$ ~\cite{DengBlote05}.
 In addition, we estimate the universal wrapping probabilities to be $R^{(x)}_{c} = 0.257\,82(6)$, $R^{(a)}_{c} = 0.459\,99(8)$, and $R^{(3)}_{c} = 0.080\,46(6)$,
 which are consistent with those estimated from bond percolation.
 In Fig.~\ref{Fig:site-R}, we show plots of $R^{(x)}$ and $R^{(a)}$ which illustrate our estimate of $p_c$.

\section{Results at $p_c$}
\label{results at pc}

 In this section, we estimate the critical exponents $y_t$, $y_h$, and $\dm$, as well as the excess cluster number.
 Fixing $p$ at our estimated thresholds for bond and site percolation, we study the covariances $g^{(x)}_{bR}$ and $g^{(x)}_{sR}$,
 the mean size of the largest cluster $C_1$, the mean size of the cluster at the origin, $\chi$, the shortest-path length $S$, and the cluster density $\rho$.
 The MC data for $g^{(x)}_{bR}$, $g^{(x)}_{sR}$, $C_1$, $\chi$ and $S$ are fitted by the ansatz
 \be
 \scrA = L^{y_\scrA} ( a_0 + b_1L^{-1.2} + b_2L^{-2})\;.
 \label{eq:A}
 \ee
 We perform fits using different combinations of the two corrections $b_1L^{-1.2}$ and $b_2L^{-2}$ and compare the results.

\subsection{Estimating $y_t$}

We estimate $y_t$ by studying the covariances $g^{(x)}_{bR}$ and $g^{(x)}_{sR}$, both of which scale as $\sim L^{y_t}$ at the critical point.
We find this procedure for estimating $y_t$ preferable to methods, such as that employed in~\cite{DengBlote05}, in which $y_t$ is estimated by studying 
how quantities behave in the neighborhood of $p_c$ as the system deviates from criticality. In particular, we believe the current method produces more reliable error estimates.

 We fit the data for $g^{(x)}_{bR}$ at $p=0.248\,811\,8$ and $g^{(x)}_{sR}$ at $p=0.311\,607\,7$ to Eq.~\eqref{eq:A}, 
 and the results are shown in Table~\ref{Tab:fit-g}.
 The estimate of $y_t$ from $g^{(x)}_{sR}$ produces a smaller error bar than that from $g^{(x)}_{bR}$.
 From these fits we take our final, somewhat conservative, estimate to be $y_t=1.141\,0(15)$.

 \begin{table}
 \caption{Fits of covariances $g^{(x)}_{bR}$ and $g^{(x)}_{sR}$.}
 \scalebox{0.95}{
 \begin{tabular}[t]{|l|l|l|l|l|l|l|} 
 \hline
                                  &$L_{\rm min}$ & $\chi^2$/DF & $y_t$ & $a_0$ & $b_1$ & $b_2$\\
 \hline
 {\multirow{4}{*}{$g^{(x)}_{bR}$}} & 16          & 4/4  & 1.140\,4(9)     & 0.231(1)   & $-0.03(2)$    & ~~0.1(2)    \\
                                   & 24          & 4/3  & 1.140\,6(13)    & 0.231(2)   & $-0.02(5)$    & ~~0.0(4)    \\
 \cline{2-7}
                                   & 16          & 4/5  & 1.140\,9(4)     & 0.230\,7(3)& $-0.012(3)$   &    ~~~~-      \\ 
                                   & 24          & 4/4  & 1.140\,6(6)     & 0.231\,1(6)& $-0.017(7)$   &    ~~~~-      \\ 
 \hline
 {\multirow{4}{*}{$g^{(x)}_{sR}$}} &  16  & 5/4  & 1.141\,6(4)     & 0.155\,1(3)& $-0.004(7)$   & $-0.06(5)$ \\
                                   &  24  & 4/3  & 1.141\,1(6)     & 0.155\,4(6)& $-0.02(2) $   & $-0.1(2) $ \\
 \cline{2-7}
                                 &  16  & 7/5  & 1.141\,1(2)     & 0.155\,5(1)& $-0.013(1) $    &    ~~~~-       \\
                                 &  24  & 4/4  & 1.141\,4(3)     & 0.155\,3(2)& $-0.010(2)$     &    ~~~~-       \\
 \hline
 \end{tabular}
  }
 \label{Tab:fit-g}
 \end{table}

 \begin{figure}
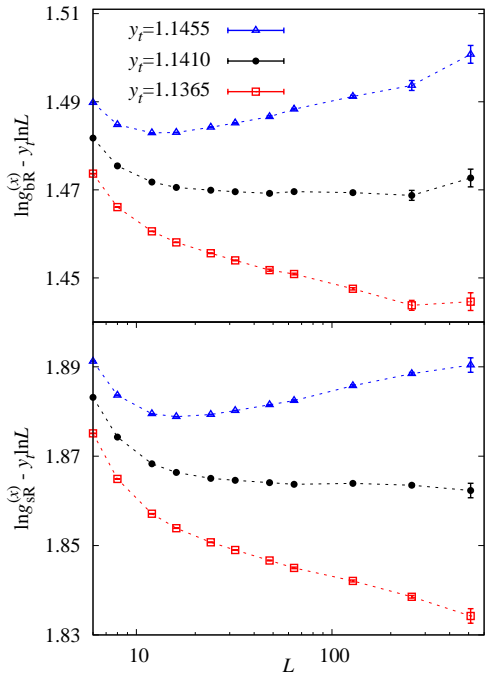

 \ytFig
 \centering
 \caption{Plots of ($\ln g^{(x)}_{bR} - y_t\ln {L}$) (top) and ($\ln g^{(x)}_{sR} - y_t\ln {L}$) (bottom) 
 vs $\ln {L}$ illustrating our estimate $y_t= 1.141\,0(15)$. The dashed curves are simply to guide the eye. }
 \label{Fig:yt}
 \end{figure}
 
 In Fig.~\ref{Fig:yt}, we plot ($\ln g^{(x)}_{bR} - y_t\ln {L}$) and ($\ln g^{(x)}_{sR} - y_t\ln {L}$) vs $\ln L$ using three different values of $y_t$:
 our estimate, as well as our estimate plus or minus three standard deviations.
 Using the true value of $y_t$ should produce a horizontal line for large $L$.
 In the figure, the data using $y_t=1.136\,5$ and $y_t=1.145\,5$ respectively bend upward and downward, suggesting that the true value of $y_t$ does indeed lie within
$3\sigma$ of our estimate. 
 The data with $y_t = 1.141$ appear to be consistent with an asymptotically horizontal line. We note that while the curve appears to be increasing around the point at $L=512$ for bond percolation, 
 it instead slightly decreases for site percolation, suggesting that in fact this movement is dominated (or even entirely caused) by noise.

 \begin{table}
 \caption{Fits of $C_1$ and $\chi$. The superscripts $b$ and $s$ denote bond and site percolation, respectively.}
  \scalebox{0.95}{
 \begin{tabular}[t]{|l|l|l|l|l|l|l|}
 \hline
                                  & $L_{\rm min}$ & $\chi^2$/DF & $y_h$ & $a_0$ & $b_1$ &  $b_2$\\
 \hline
  {\multirow{4}{*}{$C^{b}_1$}}    &   16         & 3/4  & 2.522\,86(5)    & 0.939\,4(3)   &$-0.014(6)$   & ~~$0.22(4)$  \\
                                  &   24         & 3/3  & 2.522\,89(7)    & 0.939\,3(4)   &$-0.009(11)$  & ~~$0.2(1) $  \\
 \cline{2-7}
                                  &   24         & 5/4  & 2.522\,98(3)    & 0.938\,8(2)   &~~$0.009(2)$    & ~~~~-  \\
                                  &   32         & 3/3  & 2.522\,94(4)    & 0.939\,0(2)   &~~$0.005(3)$    & ~~~~-  \\
 \hline
   {\multirow{4}{*}{$\chi^{b}$}}   &   16         & 4/4  & 2.523\,03(4)    & 1.125\,7(5)   &~~$0.14(1)$    & ~~0.18(7)   \\
                                  &   24         & 3/3  & 2.523\,00(5)    & 1.126\,2(7)   &~~$0.12(2)$    & ~~0.3(2)    \\
 \cline{2-7}
                                  &   24         & 6/4  & 2.523\,08(3)    & 1.125\,1(3)   &~~$0.157(4)$   & ~~~~-       \\
                                  &   32         & 4/3  & 2.523\,05(3)    & 1.125\,5(4)   &~~$0.151(6)$   & ~~~~-       \\
 \hline
 {\multirow{2}{*}{$C^{s}_1$}}     &    16        & 5/4  & 2.522\,99(3)      & 0.471\,16(7)  & ~~$0.024(2)$   & $-0.44(2)$  \\
                                  &    24        & 5/3  & 2.523\,00(5)      & 0.471\,1(2)   & ~~$ 0.024(4)$   & $-0.45(4)$  \\
 \hline
 {\multirow{4}{*}{$\chi^{s}$}}     &    32        & 0.9/2  & 2.522\,91(5)      & 0.284\,1(2)   & $-0.001(7)$  & $-1.15(9)$  \\
                                  &    48        & 0.7/1  & 2.522\,94(9)      & 0.284\,0(4)   & $-0.007(18)$ & $-1.3(3)$  \\
\cline{2-7}
                                  &    32        & 0.9/3  & 2.522\,92(1)      & 0.284\,06(3)  & ~~~~~~-        & $-1.16(1)$  \\
                                  &    48        & 0.9/2  & 2.522\,91(2)      & 0.284\,08(7)  & ~~~~~~-        & $-1.17(5)$  \\
\hline
 \end{tabular}
}
 \label{Tab:fit-C1}
 \end{table}

 \begin{figure}
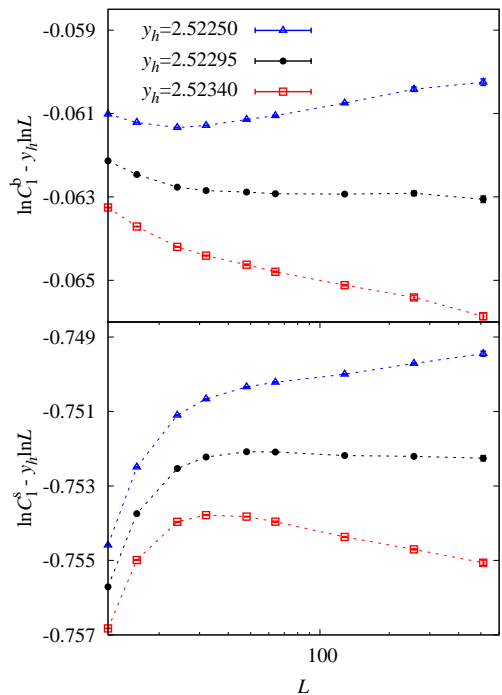

 \yhFig
 \centering
 \caption{Plots of ($\ln C^{b}_1 - y_h \ln {L}$) (top) and ($\ln C^{s}_1 - y_h \ln {L}$) (bottom) vs $\ln {L}$ to show our
          estimate $y_h = 2.522\,95(15)$. The dashed curves are simply to guide the eye.
 }
 \label{Fig:yh}
 \end{figure}
 
\subsection{Estimating $y_h$} 
 We estimate $y_h$ by studying the divergence of $C_1$ and $\chi$ as $L$ increases with $p$ fixed to our best estimates of $p_c$.
 We fit the MC data for $C_1$ and $\chi$ with Eq.~\eqref{eq:A}, 
 with the exponent $y_{\scrA}$ then corresponding to $y_h$ and $2y_h-d$, respectively.
 The results are reported in Table~\ref{Tab:fit-C1}. We use superscripts $b$ and $s$ to distinguish bond and site percolation.
 For $C^{b}_1$ and $\chi^{s}$, the amplitude $b_1$ is quite small, while $b_1$ in $\chi^{b}$ and $C^{s}_1$ is clearly present.
 In the fits of $\chi^{s}$ with one correction term $b_1L^{-1.2}$, the ratio of $\chi^2$ per DF remains large until $L_{\rm min}\ge 64$.
 We therefore show the fits with the correction $b_2L^{-2}$ instead.
 Comparing these fits, we estimate $y_h = 2.522\,95(15)$.
 
 In Fig.~\ref{Fig:yh}, we plot ($\ln C^{b}_1 - y_h \ln L$) and ($\ln C^{s}_1 - y_h \ln L$) vs $\ln L$
 using three different values of $y_h$: our estimate, as well as our estimate plus or minus three standard deviations.
 As $L$ increases, the data with $y_h = 2.522\,50$ and $2.523\,40$ respectively slope upward and downward, while 
 the data with $y_h = 2.522\,95$ are consistent with an asymptotically horizontal line.

  \begin{table}
 \caption{Fits of $S$. The superscripts $b$ and $s$ denote bond and site percolation, respectively.}
 \begin{tabular}[t]{|l|l|l|l|l|l|l|}
 \hline
                                  &$L_{\rm min}$  & $\chi^2$/DF & $\dm$            &  $a_0$       & $b_1$     & $b_2$   \\
 \hline
 {\multirow{2}{*}{${S}^{b}$}}
                                  &24             &   2/3      & 1.375\,26(5)      & 1.814\,9(5)  & ~~$-0.65(2)$    &   $-3.8(2)$    \\
                                  &32             &   1/2      & 1.375\,33(7)      & 1.814\,2(7)  & ~~$-0.59(5)$    &   $-4.4(4)$    \\
                                  &48             &   0/2      & 1.375\,30(9)      & 1.815(1)     & ~~$-0.63(9)$    &   $-4(1)$    \\
 \hline
 {\multirow{3}{*}{${S}^{s}$}}
                                  &16             &  5/4       & 1.375\,80(2)      & 1.383\,4(2)  & ~~$-3.432(5)$   &   $2.72(3)$    \\
                                  &24             &  4/4       & 1.375\,77(3)      & 1.383\,6(3)  & ~~$-3.45(2)$    &   $2.82(3)$    \\
                                  &32             &  4/2       & 1.375\,76(5)      & 1.383\,7(4)  & ~~$-3.45(3)$    &   $2.9(3)$    \\
 \hline
 \end{tabular}
 \label{Tab:fit-S}
 \end{table}

 \begin{table}
 \caption{Fits of $\rho$. The superscripts $b$ and $s$ denote bond and site percolation, respectively.}
 \scalebox{1.00}{
 \begin{tabular}[t]{|l|l|l|l|l|l|}
 \hline
                                  &$L_{\rm min}$  & $\chi^2$/DF & $\rho_{c}$      &  $b$       & $b_1$\\
 \hline
  {\multirow{4}{*}{${\rho}^{b}$}}   &16             & 3/5    & 0.272\,932\,83(1) & 0.679(3)   & ~~$0.1(6)$  \\
                                    &24             & 1/4    & 0.272\,932\,83(1) & 0.674(6)   & ~~$3(4)  $  \\
 \cline{2-6}
                                    &16             & 2/7    & 0.272\,932\,83(1) & 0.678\,9(6)& ~~~~-  \\
                                    &24             & 2/6    & 0.272\,932\,83(1) & 0.679(2)   & ~~~~-  \\

 \hline
  {\multirow{4}{*}{${\rho}^{s}$}}   &12             & 4/6    & 0.052\,438\,218(3)& 0.674\,5(5) & ~~$ 0.02(8)$ \\
                                    &16             & 4/5    & 0.052\,438\,218(3)& 0.674\,7(8) & $-0.02(21)$ \\
                                    &24             & 4/4    & 0.052\,438\,218(3)& 0.674(2)    & ~~$0.2(10)$ \\
\cline{2-6}
                                    &12             & 4/7    & 0.052\,438\,218(3)& 0.674\,6(2) & ~~~~-  \\
                                    &16             & 4/6    & 0.052\,438\,218(3)& 0.674\,6(3) & ~~~~-  \\
                                    &24             & 4/5    & 0.052\,438\,218(3)& 0.674\,6(5) & ~~~~-  \\
 \hline
 \end{tabular}
 }
 \label{Tab:fit-rhok}
 \end{table}

\begingroup
\squeezetable
\begin{table*}
  \begin{center}
  \caption{Summary of estimated thresholds, critical exponents, universal wrapping probabilities, and excess cluster number of bond 
   and site percolation on the simple-cubic lattice.
 We note that the values of $y_t$ and $y_h$ in~\cite{DengBlote05} marked by the superscript $\ast$ contained typographical errors.
 The final error bars reported in~\cite{DengBlote05} were also underestimated, taking insufficient account of systematic errors.} 
  \begin{tabular}[t]{llllllllllll} 
    \hline
    \hline
    Ref.                         &     $p_c$(bond)        &   $p_c$(site)      &$y_t=1/\nu$&$y_h=d_f$ & $\dm$ &    $y_i$& $R^{(x)}$ &  $R^{(a)}$&$R^{(3)}$  &   $b$      \\
    \hline
    \cite{LorenzZiff98a}        &    $0.248\,812\,6(5)$  &                    &   1.12(2)    & 2.523(4)     &       &         &              &      &&      \\
    \cite{LorenzZiff98b}        &                        & $0.311\,608\,0(4)$ &              &              &       &         &              &      &&      \\
    \cite{BallesterosFernandezMayorSudupeParisi99}  &     & $0.311\,608\,1(13)$& $1.141(2)$   & $2.523\,0(3)$&       &$-1.61(13)$&            &      &&      \\
    \cite{MartinsPlascak03}      &    $ 0.249\,0(2)$      & $0.311\,5(3)$      & $1.15(2)$    &              &       &         &  0.265(6)    &0.471(8)   &0.084(4)  &  \\
    \cite{DengBlote05}           &                        & $0.311\,607\,7(4)$ &   1.145\,0(7)& 2.522\,6(1) &       &          &               &    &&\\
    \cite{ZhouYangZiffDeng12}   &    0.248\,812\,0(5)    &                    &   1.142(3)   & 2.523\,5(8)  &       &         &              &      &&       \\
    \cite{ZhouYangDengZiff12}   &                        &                    &              &              &1.375\,6(6)&       &              &      &&       \\
    \cite{KozlovLagues10}        &                        &                    &  1.142(8)    &              &       &$-1.0(2)$&              &      &&          \\
    This work                    &    0.248\,811\,82(10)  &0.311\,607\,7(2) &1.141\,0(15) & 2.522\,95(15)&1.375\,6(3)&$-1.2(2)$& 0.257\,80(6) &0.459\,98(8)&0.080\,44(8)& $0.675(2)$  \\
    \hline
  \end{tabular}
  \label{Tab:summary}
  \end{center}
\end{table*}
\endgroup
\subsection{Estimating $\dm$}
 We estimate the shortest-path fractal dimension $\dm$ by studying the quantity $S$ at our estimated thresholds.
 The MC data for $S$ are fitted to Eq.~\eqref{eq:A} with the exponent $y_\scrA$ replaced by $\dm$, and the results are reported in Table~\ref{Tab:fit-S}.
 We again use the superscripts $b$ and $s$ to distinguish bond and site percolation.
 In the fits, both $b_1$ and $b_2$ are clearly observable for $S^{b}$ and $S^{s}$.
 And when we set $b_2=0$, the ratio of $\chi^2$ per DF remains relatively large. 
 We also did the fits by replacing the correction with $b_2$ by a constant term $c_0$ in Eq.~\eqref{eq:A}, 
 and obtained $\dm ({\rm bond}) = 1.375 \, 55 (6)$ and $\dm ({\rm site}) = 1.375 \, 59 (6)$.
 Comparing these fits, we estimate $\dm = 1.375\,6(3)$.

\begin{figure}
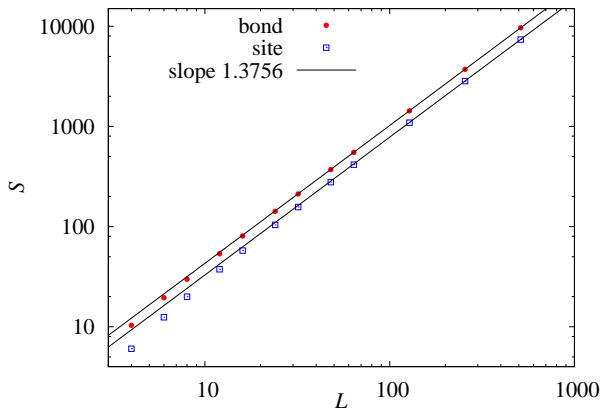

 \dminFig
 \centering
 \caption{Log-log plot of $S$ versus $L$ for bond and site percolation. Two straight lines with slope 1.375\,6 are included for comparison.
 }
 \label{Fig:dmin}
 \end{figure}

 To illustrate this estimate, Fig.~\ref{Fig:dmin} shows a log-log plot of $S$ versus $L$.

 \subsection{Excess number of clusters}
 The cluster density tends to a finite limit $\rho_c=\lim_{L\to\infty}\lim_{p\to p_c}\rho$ at criticality.
 While the value of $\rho_{c}$ is non-universal, the {\em excess cluster number}
 $b := \lim_{L\to \infty}\lim_{p\to p_c} L^d(\rho - \rho_{c})$ is universal \cite{ZiffFinchAdamchik97}.
 To estimate $b$, we study $\rho$ with $p$ fixed to our estimated thresholds for bond and site percolation and fit the data to the ansatz
 \be
 \rho = \rho_{c} + L^{-3}(b + b_1L^{-2}) \; .
 \label{eq:rhok}
 \ee
 
 The resulting fits are summarized in Table~\ref{Tab:fit-rhok}, where we again use superscripts $b$ and $s$ to differentiate the bond and site cases.
 We report fits both with $b_1$ free and with $b_1=0$.
 We find that $\rho$ can be well fitted to~\eqref{eq:rhok} with $b_1=0$ fixed. 
 Leaving $b_1$ free, we find that $b_1$ is consistent with zero, suggesting that the leading correction exponent might be even smaller than $-2$.
 We also performed fits in which the leading correction exponent was fixed to $-1.2$ and $-3$, and in both cases the resulting estimates of $\rho_c$ and $b$ were consistent
 with those reported in Table~\ref{Tab:fit-rhok}.
 Leaving the leading correction exponent free produces unstable fits however.
 Comparing these fits, we estimate $b = 0.675(2)$.
 
 Our estimate of $b$ is determined on th periodic $L\times L\times L$ simple cubic lattice; on the $L\times L$ square lattice $b=0.883\,5(8)$ \cite{ZiffFinchAdamchik97}. 
 The excess cluster number was studied in \cite{LorenzZiff98a} on an $L\times L\times L'$ lattice with $L'\gg L$.
 Naively, extrapolating their results to $L'=L$ gives an estimate of $b\approx 0.412$ which is significantly below our estimate.
 We also note that our estimate of the number of clusters $\rho_c=0.272\,932\,83(1)$ differs slightly from the estimate $\rho_c=0.272\,931\,0(5)$ reported in \cite{LorenzZiff98a}.

\section{Discussion}
\label{discussion}
 We study in this paper standard bond and site percolation on the three-dimensional simple-cubic lattice with periodic boundary conditions.
 Using extensive Monte Carlo simulations and finite-size scaling analysis, 
 we report the estimates: $p_c = 0.248\,811\,82(10)$ (bond) and $p_c = 0.311\,607\,7(2)$ (site).
 The bulk thermal and magnetic exponents are estimated to be $y_t = 1.141\,0(15)$ and $y_h = 2.522\,95(15)$, 
 the shortest-path fractal dimension to be $\dm=1.375\,6(3)$, and the leading irrelevant exponent to be $y_i=-1.2(2)$.
 The universal value of the excess cluster number is estimated to be $b=0.675(2)$.

 We emphasize that the reported estimates of $p_c$ are obtained by studying wrapping probabilities,
 which are found to have weaker corrections to scaling than dimensionless ratios constructed from moments 
 of magnetic quantities such as $\scrC_1$ and $\scrS_m$.
 In particular, we find evidence suggesting that the leading correction exponent in certain wrapping 
 probabilities ($R^{(x)}$ and $R^{(a)}$ for bond percolation, $R^{(3)}$ for site percolation) may be $\approx -2$ rather than $-1.2$, although the reasons are not clear.
 The universal values of the wrapping probabilities 
 we studied are estimated to be: $R^{(x)}_{c} = 0.257\,80(6)$, $R^{(a)}_{c} = 0.459\,98(8)$, and $R^{(3)}_{c} = 0.080\,44(8)$,
 by comparing the results for bond and site percolation.

 From these values we can estimate other wrapping probabilities discussed in the literature, such as 
 \begin{equation}
 \begin{aligned}
 \! R^{(1)}  : &= \langle \scrR^{(x)}(1-\scrR^{(y)})(1-\scrR^{(z)}) \rangle \nonumber \\ 
               &= \frac{1}{3}(2R^{(a)} + R^{(3)} - 3R^{(x)})\;, \nonumber \\
 \! R^{(2)}  : &= \langle \scrR^{(x)}\scrR^{(y)}(1 - \scrR^{(z)})\rangle   \nonumber  \\
               &= \frac{1}{3}(3R^{(x)} - 2R^{(3)} - R^{(a)})\;, \nonumber \\
 \! R^{(x,y)} :&= \langle \scrR^{(x)}\scrR^{(y)} \rangle = \frac{1}{3}(3R^{(x)} +  R^{(3)} - R^{(a)}) \;.
 \end{aligned}
 \end{equation}
 In words, $R^{(1)}$ is the probability that a winding exists in one given direction but not in the other two directions;
 $R^{(2)}$ is the probability that a winding exists in two given directions but not in the third;
 and $R^{(x,y)}$ is the probability that a winding exists in two given directions, regardless of whether a winding exists in the third direction.
 We obtain $R^{(1)}_{c} = 0.075\,67(14)$, $R^{(2)}_{c} = 0.050\,85(14)$, and $R^{(x,y)}_{c} = 0.131\,29(12)$.

 Table~\ref{Tab:summary} summarizes the estimates presented in this work.
 For comparison, we also provide an (incomplete) summary of previous estimates.

\section{Acknowledgments}
 This research was supported in part by NSFC under GrantS No. 91024026 and No. 11275185, and the Chinese Academy of Science.
 It was also supported under the Australian Research Council's (ARC) Discovery Projects 
 funding scheme (Project No. DP110101141), and T.G. acknowledges support from the Australian Research Council through a Future Fellowship
(Project No. FT100100494).
 Y.D. is grateful for the hospitality of Monash University at which this work was partly completed. 
 The simulations were carried out on the NYU-ITS cluster, which is partly supported by NSF Grant No. PHY-0424082.

%\bibliography{bib.bib}

\end{document}